\documentclass[fleqn,usenatbib]{mnras}

\usepackage{newtxtext,newtxmath}
\usepackage[T1]{fontenc}
\DeclareRobustCommand{\VAN}[3]{#2}
\let\VANthebibliography\thebibliography
\def\thebibliography{\DeclareRobustCommand{\VAN}[3]{##3}\VANthebibliography}

\usepackage{graphicx}	
\usepackage{amsmath}	

\title[Search for transients with Arecibo]{Search for fast radio transients using Arecibo drift-scan observations at 1.4 GHz}

\author[Perera et al.]{B.~B.~P.~Perera,$^1$\thanks{E-mail: bhakthiperera@gmail.com} A.~J.~Smith,$^1$ S.~Vaddi,$^1$ R.~Carballo-Rubio,$^2$ A.~McGilvray,$^1$ A.~ Venkataraman,$^1$ 
\newauthor D.~Anish~Roshi,$^1$ P.~K.~Manoharan,$^1$ P.~Perillat,$^1$ E.~Lieb,$^3$ 
D.~R.~Lorimer,$^{4,5}$ M.~A.~McLaughlin,$^{4,5}$ \newauthor
D.~Agarwal,$^{4,5}$ K.~Aggarwal$^{4,5}$ and S.~M.~Ransom$^6$\\
\\ $^1$ Arecibo Observatory, University of Central Florida, HC3 Box 53995, Arecibo, PR 00612, USA
\\ $^2$ Florida Space Institute, University of Central Florida, 12354 Research Parkway, Partnership 1, Orlando, FL 32826, USA 
\\ $^3$ Department of Astrophysical \& Planetary Sciences, University of Colorado, Boulder, CO 80309, USA
\\ $^4$ Department of Physics and Astronomy, West Virginia University, Morgantown, WV 26501, USA
\\ $^5$ Center for Gravitational Waves and Cosmology, West Virginia University, Chestnut Ridge Research Building, Morgantown, WV 26505, USA
\\ $^6$ National Radio Astronomy Observatory, 520 Edgemont Rd., Charlottesville, VA 22903, USA
}

\date{Accepted XXX. Received YYY; in original form ZZZ}

\pubyear{2021}

\begin{document}
\label{firstpage}
\pagerange{\pageref{firstpage}--\pageref{lastpage}}
\maketitle

\begin{abstract}
We conducted a drift-scan observation campaign using the 305-m Arecibo telescope in January and March 2020 when the observatory was temporarily closed during the intense earthquakes and the initial outbreak of the COVID-19 pandemic, respectively. The primary objective of the survey was to search for fast radio transients, including Fast Radio Bursts (FRBs) and Rotating Radio Transients (RRATs). We used the 7-beam ALFA receiver to observe different sections of the sky within the declination region $\sim$(10--20)~deg on 23 nights and collected 160 hours of data in total. We searched our data for single-pulse transients, covering up to a maximum dispersion measure of 11\,000~pc~cm$^{-3}$ at which the dispersion delay across the entire bandwidth is equal to the 13~s transit length of our observations. The analysis produced more than 18 million candidates. Machine learning techniques sorted the radio frequency interference and possibly astrophysical candidates, allowing us to visually inspect and confirm the candidate transients. We found no evidence for new astrophysical transients in our data. We also searched for emission from repeated transient signals, but found no evidence for such sources. We detected single pulses from two known pulsars in our observations and their measured flux densities are consistent with the expected values. Based on our observations and sensitivity, we estimated the upper limit for the FRB rate to be $<$2.8$\times10^5$~sky$^{-1}$~day$^{-1}$ above a fluence of 0.16~Jy~ms at 1.4~GHz, which is consistent with the rates from other telescopes and surveys.
\end{abstract}

\begin{keywords}
transients: fast radio bursts -- stars: neutron -- pulsars: general -- pulsars: individual: PSR J0625$+$10, J1627$+$1419
\end{keywords}



\section{Introduction}

FRBs are millisecond-duration energetic radio pulses that are accompanied by large dispersion measures (DMs, where DM  is the integral of the column density of free electrons along the line-of-sight from the Earth to the source), placing their origin at cosmological distances. The majority of initial FRBs were discovered with the Parkes radio telescope \citep{lbm+07, tsb+13} and since then, more than 600 FRBs have been detected using various other telescopes\footnote{\url{https://www.frbcat.org}}$^,$\footnote{\url{https://www.chime-frb.ca/catalog}}, covering observing centre frequencies below 1.4~GHz \citep{sch+14,mls+15,cfb+17,bsm+17,smb+18,abb+19a,zll+20,lbp+20,rcd+19,cvo+20}, and including some follow-up observations at 4--8~GHz \citep{msh+18,gsp+18}. The DMs of  FRBs \citep[currently in the range $\sim$100--3000~pc~cm$^{-3}$; ][]{aab+21} are much larger than the expected DM contributions from the Milky Way in the direction of their lines-of-sight, indicating that the origins must be extra-galactic. The FRBs were identified as one-time transient events until the first repeating emission was detected with the Arecibo telescope from FRB121102 \citep{ssh+16}. Since then, more than 20 FRBs have been identified as repeaters \citep{lwm+20,abb+19b,abb+19c,kso+19}. Using follow-up observations with the VLA, FRB121102 was localized to a star-forming region in a dwarf galaxy at a redshift $z$ of 0.193 \citep{clw+17}, providing further evidence for FRBs that are cosmological in origin. Since then, 17 additional FRBs have been localized to sub-arcsecond precision by radio interferometers, associating them with host galaxies\footnote{\url{http://frbhosts.org}} with known redshifts $z\sim0.034-0.66$ \citep[e.g.,][]{lbp+20, mnh+20, mpm+20, bdp+19, pmm+19, rcd+19, clw+17,hps+20,pbt+21,bha+21}.

While we have evidence that FRBs are associated with host galaxies, these galaxies have a range of properties, and the FRB origins and formation mechanisms are still a mystery.
Determining the prevalence of FRBs across cosmic time could provide significant insight into their formation mechanisms.  The Galactic magnetar SGR J1935$+$2154 emitted FRB-like bursts during its recent active phase \citep{abb+20,brb+20}. These bursts are highly energetic, brighter than any radio bursts seen from Galactic sources, and only a few orders of magnitude lower than the equivalent energy from the faintest FRB \citep{bbb+21}. This leads to the possibility that at least some FRBs are associated with magnetars \citep[see][]{lp20}. In order to identify a possible neutron star (NS) origin for FRBs, periodic searches (with periods of order milliseconds to seconds) have been carried out, but these were not successful \citep{zgf+18}. However, long periodicities of $\sim$16 days and $\sim$157 days have been reported for FRB180916B \citep{aab+20} and FRB121102 \citep{rms+20,css+21}, respectively, suggesting that FRB emission mechanisms perhaps align with putative binary systems \citep{iz20,lbg20}, slowly spinning neutron stars \citep{bwm20}, and spin precession, including orbit-induced precession \citep{yz20}, forced precession \citep{sob20}, and free precession models \citep{lbb20}. Based on the properties of host galaxies, FRBs may also originate through superluminous supernovae and long gamma-ray bursts \citep[for reviews, see][]{xwd+21,zha20a}. Moreover, some suggest that FRBs are  produced through the magnetospheric activities of magnetars that have been newly formed through binary NS mergers or pre-merger NS-NS interactions \citep{zha20b,wwy+20}. However, none of these mechanisms are exclusively capable of explaining the origin of FRBs in general. 

Recently, \citet{pag+21} carried out a search  targeting FRBs associated with a sample of 11 gamma-ray bursts that show evidence for the birth of a magnetar. However, this study was unsuccessful in finding FRB-like signals associated with these sources. To uncover the physics of these phenomena, it is essential to increase the known FRB population, to measure a broad range of characteristics such as their duration, emission component structure, spectral properties, polarization, and to also achieve localization in order to measure the properties of the host galaxy. FRBs are detected through single-pulse search algorithms due to their non-periodic, energetic, short-duration, and single-event nature. Many telescopes utilize real-time FRB detection instruments with single-pulse search pipelines to improve the speed and the efficiency of new discoveries, enabling the observations to be done commensally with other projects and ultimately maximizing telescope time for FRB searches and characterization studies \citep[e.g., ][]{abb+18a,kca+15,fkg+18, sal+19, aab+21,nll+21}.

\begin{figure}
\includegraphics[width=8.6cm]{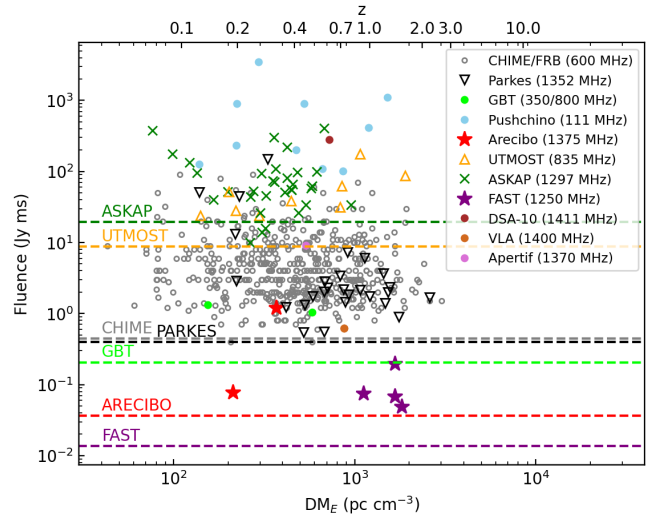}
\caption{
Excess DM and fluence for all known FRBs to-date. The excess DM is obtained after subtracting the expected Galactic contribution based on the electron density model YMW16 \citep{ymw17} along the line-of-sight of the source. The upper x-axis represents the redshift and is calculated using the method given in \citet{iok03}. More details are given in Section~\ref{pipeline}. The horizontal {\it dashed} lines represent the single-pulse sensitivity for some of the telescopes, assuming a pulse width of 1~ms and a flat spectral index (see Section~\ref{sensitivity}). The typical center frequencies of these FRB detections and surveys are given in parentheses.
}
\label{fluence}
\end{figure}

FRBs with high DMs are potentially important as they are generally accompanied by high redshifts and provide a unique opportunity to probe the far reaches of the intergalactic medium. This is crucial for constraining the epochs of hydrogen and helium reionization produced by the ignition of the earliest stars and galaxies \citep[see][]{khcp15}, leading to a better understanding of the universe in general. The largest DM observed so far is $3038.06\pm0.02$~pc~cm$^{-3}$ for FRB20180906B\footnote{\url{https://www.chime-frb.ca/catalog}}, placing its origin at a redshift $>$2 \citep{aab+21}. Some distant FRBs with large DMs exhibit low fluences, indicating the importance of highly sensitive telescopes in detecting those weak sources \citep{2018NatAs...2..860L,zha18}. The highly sensitive 500-m FAST telescope has discovered four FRBs so far, all of which are accompanied by high DMs ($>$1000~pc~cm$^{-3}$) and low fluences ($<$0.2~Jy~ms) -- e.g., FRB 181017.J0036+11 has a DM of $1845.2\pm1$~pc~cm$^{-3}$ and a fluence of 0.042 Jy ms, which is the faintest FRB detected so far \citep{nll+21,zll+20}. Moreover, both FRBs detected by the Arecibo telescope have low fluences -- 0.08 and 1.2~Jy~ms for FRB141113 and FRB121102, respectively \citep{sch+14, pab+18} -- and are in the lower end of the fluence distribution of known FRBs (see Fig.~\ref{fluence}). In general, detecting FRBs with high redshift and low fluence is challenging due to sensitivity limitations of telescopes. Hence, large aperture instruments such as the FAST and Arecibo telescopes are of paramount importance for detecting such weak sources \citep[see][]{zha18}. However, the small field-of-view of such large telescope degrades the speed of the surveys.

RRATs are rotating neutron stars that emit sporadic emission that can be discovered only through single-pulse search algorithms \citep{mll+06}. They have much more similar emission properties to normal pulsars than FRBs. Unlike FRBs, RRATs are located within our Galaxy\footnote{\url{http://astro.phys.wvu.edu/rratalog}}. Utilizing a large DM range in single-pulse search algorithms covering the Galactic DMs, we can detect both of these fast transient sources, in addition to normal or giant pulse emitting pulsars, and this is a standard approach in many surveys \citep[see][for details]{pab+18,pck+20}.

In this work, we carried out a drift-scan observation campaign at the Arecibo Observatory to search for fast radio transients, including FRBs and RRATs. The paper is organized as follows: we describe our observations, data preparation, and the system performance in Section~\ref{obs}. The sensitivity of our data to single pulses is estimated in Section~\ref{sensitivity} and we discuss our search pipeline in Section~\ref{pipeline}. The results of our single-pulse search are presented in Section~\ref{results}, and the detection of known pulsars in our data and their flux densities are described in Section~\ref{known}. Finally in Section~\ref{dis}, we discuss results and limits on FRB rates based on our observations.

\section{Observations}
\label{obs}

\begin{figure}
\includegraphics[width=8.5cm]{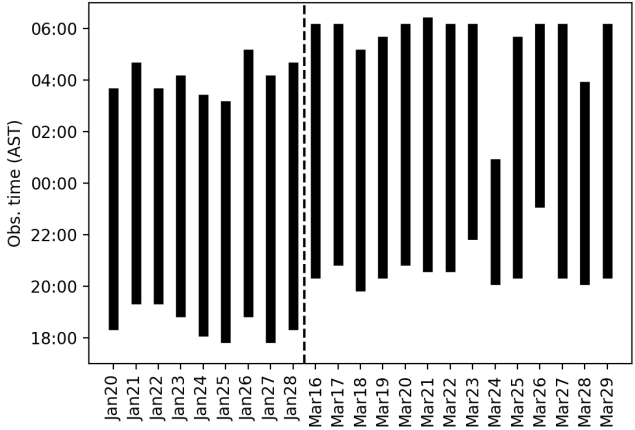}
\caption{
The observation program of the drift-scan campaign. We observed 9 and 14 days consecutively in January and March, respectively.
Each day, we started the observation between 18:00--21:00 AST, except on March 26 where we started at 23:15 AST due to technical difficulties. We observed for 8--10 hours each day except only 4.5 hours on March 24 due to data storage issues.
}
\label{schedule}
\end{figure}

A series of strong earthquakes struck Puerto Rico starting on 28 December 2019, including a magnitude of 6.0 catastrophic earthquake\footnote{\url{https://www.usgs.gov/news/magnitude-64-earthquake-puerto-rico}} on 6 January 2020 in the southwest of the island. Following these events, the Arecibo Observatory was temporarily closed for several weeks in January 2020 for safety inspections of the facility. Shortly thereafter, the COVID-19 pandemic also forced a temporary closure of the observatory site during March 2020. During both of these periods, we were able to coordinate and conduct drift-scan observations with the 305-m Arecibo telescope with minimum operational support by recording the data continuously as the sky drifted across the telescope beam at the sidereal rate. We observed everyday between 20--28 January and 16--29 March, resulting in 23 days in total. Each day, we started observations around 20:00 AST (Atlantic Standard Time) and continued taking data for about 8--10 hours. The observation program and the time duration on each day are presented in Fig.~\ref{schedule}.
We further note that this campaign was conducted during the downtime of the telescope until its normal operations resumed. Therefore, we could not collect more data after this period.

The primary goal of our campaign was to search for new transients in the Arecibo sky. The transient searches typically require high time and frequency resolution. FRBs were mainly detected at $\sim$1.4~GHz frequencies until the recent CHIME telescope detections at $\sim$600~MHz\footnote{\url{https://www.chime-frb.ca/catalog}} \citep{aab+21}. We carried out our observations using the Arecibo L-band Feed Array\footnote{\url{http://www.naic.edu/alfa/}} (ALFA) receiver with a centre frequency of 1375~MHz and a bandwidth of 322~MHz with 960 frequency channels, resulting in a spectral channel size of 0.335~MHz. The two polarization channels were summed together, and we recorded the total intensity data using the Mock spectrometers at a sampling rate of 65~$\mu$s. We note that this setup is very similar to the regular observing configuration of the PALFA survey\footnote{\url{http://www2.naic.edu/alfa/pulsar}}, which has discovered several fast transients\footnote{\url{http://www.naic.edu/~palfa/newpulsars}}, including  both Arecibo-discovered FRBs and also 20 RRATs \citep{sch+14,pab+18,dcm+09,psf+21}.

The field-of-view (FoV) is important in regular searching campaigns as it can enhance the instantaneous sky coverage and thus, the survey speed. The ALFA receiver was a seven-beam feed-array which observed seven pixels on the sky simultaneously. Each beam had a full-width-half-maximum (FWHM) power of approximately $3'.35$, corresponding to an instantaneous FoV of 0.022~deg$^2$ across the 7 pixels \citep[see][]{cfl+06,sch+14}. With this FWHM beamwidth, the sky transit time between the half power points of the beam was approximately 13 ~s. Therefore, the ALFA receiver in drift-scan mode provided rapid sky coverage.

\begin{figure}
\includegraphics[width=8.5cm]{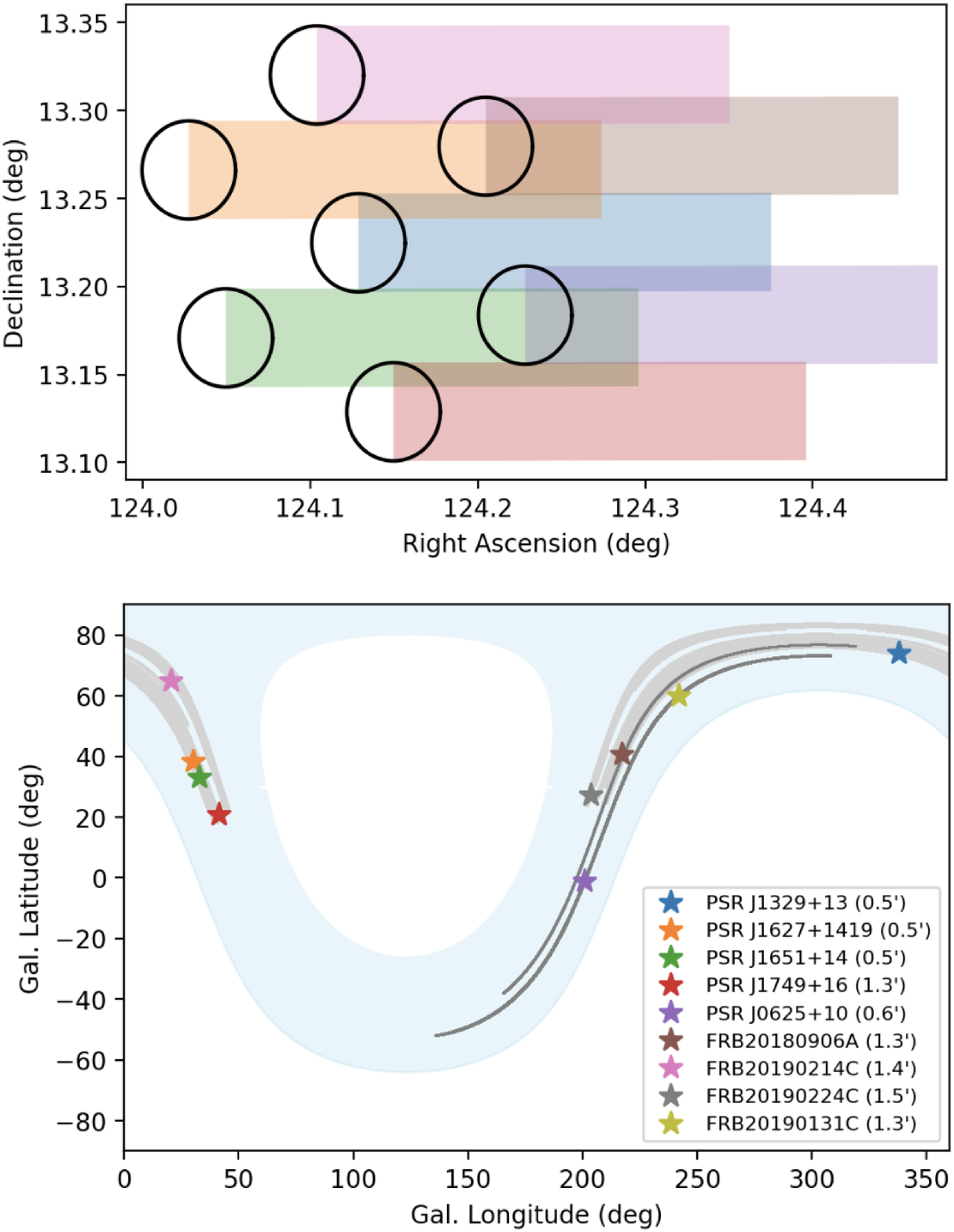}
\caption{
{\it Top:} The circles represent the position and the FWHM power of the ALFA receiver beams with respect to the sky at the beginning of the 16 March observation. The color stripes represent the first 60~s of the observation, showing no gaps across the sky between the beams during the scan.
{\it Bottom:} The sky regions covered during our drift-scan observations displayed in Galactic coordinates. The {\it blue} shaded area represents the portion of the sky visible to the Arecibo telescope within its declination range from $-1^\circ$ and $+37.5^\circ$. The coverage of January and March observations are marked in {\it dark gray} and {\it light gray}, respectively. The known pulsars and FRBs within the ALFA beams are marked in different colors, and the offsets from the centre of the closest observed beam are given in parentheses in units of arc-minutes (see Section~\ref{known} for details of the detection of these sources). 
}
\label{sky}
\end{figure}

With the limited operational support at the observatory in January, we were not able to move the receiver in azimuth and zenith angles. Rather we kept it at the same position (with the exception of January 28), resulting in observing at the same zenith angle. During the March observations, we were able to change the receiver position in azimuth by one degree each day 
in order to scan different parts of the sky. 
In addition, the receiver was positioned at a parallactic angle offset of 19$^\circ$ in all our observations to scan the sky without leaving gaps between the beams (see Fig~\ref{sky} top panel) and then unchanged during data-taking.
Fig.~\ref{sky} bottom panel shows the observed sections of the sky during the two observing sequences. Several known pulsars and FRBs happened to be within the ALFA beams, and they are marked in the figure. We discuss the detectability of these sources in Section~\ref{known}.

\subsection{Data preparation}

The data were recorded in PSRfits format\footnote{\url{https://www.atnf.csiro.au/research/pulsar/psrfits_definition/Psrfits.html}} as 16-bit integers, and the lower and upper half of the frequency bands were recorded separately as two individual files. Since our single-pulse searching software supports only 8-bit integers, we compressed the original 16-bit data into 8-bit using \texttt{psrfits2psrfits}\footnote{\url{https://github.com/juliadeneva/psrfits2psrfits}} and then merged the two frequency bands using \texttt{combine\_mocks}\footnote{\url{https://github.com/demorest/psrfits_utils}} to produce the full bandwidth of 322~MHz. As required by our single-pulse search pipeline, the data were then converted into the filterbank format using \texttt{digifil}\footnote{\url{http://dspsr.sourceforge.net/index.shtml}} \citep{vb11}. All these tools are commonly used in pulsar data preparation, handling, and processing.

\begin{figure}
\includegraphics[width=8.6cm]{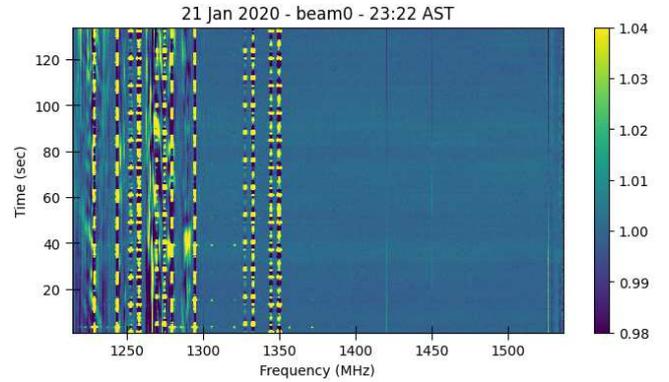}
\caption{
The dynamic spectrum of approximately 130~s of data obtained on 21 January 2020 at 23:22 AST. A high level of RFI can be seen below 1300~MHz in general and also around 1330 and 1350~MHz. We excised the spectral channels corresponding to these frequency ranges in the data processing to mitigate RFI.
}
\label{dynamic}
\end{figure}

We investigated the radio frequency interference (RFI) environment across the frequency band. By inspecting the dynamic spectra of the data (see Fig.~\ref{dynamic} for example), we identified persistent RFI in a large number of frequency channels between 1250 and 1300~MHz. In addition, the level of the RFI below 1250~MHz was dynamic and varied over our observation time. Thus, we decided to ignore and excise the frequency channels below 1300~MHz across all our observations. We also identified strong RFI around 1330 and 1350~MHz and excised the relevant channels. The RFI removal reduced the usable bandwidth by roughly one-third, resulting in an effective bandwidth of approximately 215~MHz.

\subsection{System gain and stability}
\label{stability}

\begin{figure}
\includegraphics[width=8.3cm]{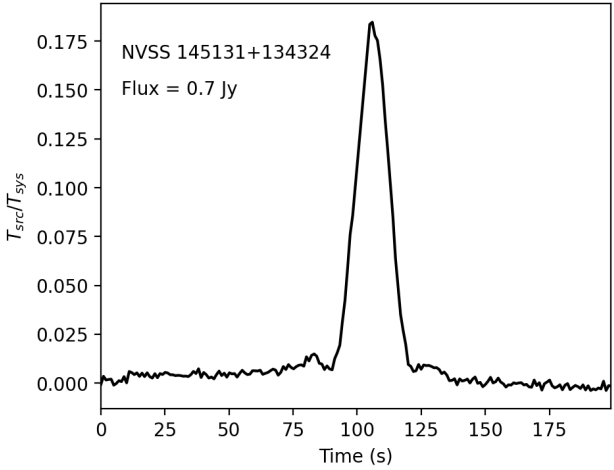}
\caption{
The detection of continuum background source NVSS 145131$+$134324 on March 17 in the central beam. The y-axis is scaled as the ratio of the source temperature and the system temperature. The source has a known flux density of 692~mJy at 1.4~GHz and was used to estimate the gain of the telescope as described in Section~\ref{stability}. 
}
\label{nvss}
\end{figure}

We detected radio continuum emission from many background sources during their transit across the ALFA beams. Based on their sky locations, we confirm that these are identified radio point sources (i.e., source size is smaller than our beam size) reported in the NVSS survey catalog  \footnote{\url{https://www.cv.nrao.edu/nvss}} \citep{ccg+98}. These compact continuum sources were detected in almost all beams daily in our observations.  Fig.~\ref{nvss} shows the detection of NVSS 145131$+$134324 on March 17 in the central beam (Beam0) of the receiver. We used these detected sources along with their flux densities reported in the NVSS catalog to estimate the gain of the telescope and the stability of the system throughout our observations. The details of the gain estimation method are described in Appendix~\ref{gain}. 
We estimated the gain for all beams (except Beam5) for most observing sessions. We ignored Beam5 in the gain calculation as one of its polarization channels was unstable and poorly behaved. However, we note that Beam5 was included in the single-pulse search analysis described in Section~\ref{pipeline}. Our system performance analysis determined that the overall system was stable throughout the campaign, while showing a slight difference in gain between the January and March observing sessions. The average gain of the central beam, Beam0, was estimated to be $8.2\pm0.5$~K~Jy$^{-1}$ and that of the other beams was $7.1\pm0.8$~K~Jy$^{-1}$ in March observations, which is consistent with the typical system performance of the ALFA receiver\footnote{\url{http://www.naic.edu/\%7Ephil/mbeam/mbeam.html}}. In January, the average gain of Beam0 and the other beams were $7.5\pm0.3$ and $5.6\pm0.8$~K~Jy$^{-1}$, respectively. We attribute the lower gain to the inactivity of the tie-down cables of the telescope throughout the January observations, which resulted in defocusing compared to the optimal telescope setup. However, we note that even during this period the system sensitivity was still very high compared to other telescopes in the world, with the exception of the 500-m FAST telescope.

\section{Sensitivity to single pulses}
\label{sensitivity}
From radiometer noise considerations, the peak flux density of a single pulse,
\begin{equation}
\label{s_peak}
S_i = \frac{\beta (T_{sys}+T_{sky}) (S/N)_b}{GW_i}\sqrt{\frac{W_b}{n_p \Delta f}},
\end{equation} 
where $T_{\rm sys}$ is the system temperature at the observing frequency, $T_{\rm sky}$ is the sky temperature in the direction of the telescope pointing, $\beta$ is the factor of sensitivity loss due to digitization, $(S/N)_{\rm b}$ is the signal-to-noise ratio of the broadened pulse, $G$ is the telescope gain, $\Delta f$ is the observing bandwidth, $n_p$ is the number of summed polarization channels, $W_i$ is the intrinsic pulse width, and $W_b$ is the broadened pulse width \citep[see][]{cm03,pab+18}. The pulse can be broadened due to various reasons such as dispersion smearing within the frequency channel, scattering, etc. Assuming a telescope gain of $8$~K~Jy$^{-1}$ (see Section~\ref{stability}) and typical values for the system $T_{\rm sys} + T_{\rm sky} \approx 30$~K, $\Delta f = 215$~MHz, $n_p = 2$, and $\beta = 1.1$ \citep{pab+18}, we can rewrite the above expression as $S_i = 0.006 (S/N)/\sqrt{W}$~Jy for pulse width $W_i = W_b = W$ in units of milliseconds.  

The above expression gives a theoretical estimate of the single-pulse sensitivity assuming Gaussian noise in the data. However, the sensitivity in real data can deviate significantly from this formalism due primarily to the presence of RFI and non-Gaussian noise. Based on simulations using PALFA data, \citet{pab+18} found that the sensitivity of their pipeline to   single pulses was degraded by many factors depending on the dispersion measure and the pulse width of the detection. They determined a degradation factor of approximately 1.5 for a pulse with a DM of 1000~pc~cm$^{-3}$ and a width of 5~ms. Since our observation configuration is similar to that of PALFA, we assume the same degradation factor in our study as a conservative value. Assuming a $S/N=8$ and $W=5$~ms, the single-pulse sensitivity in our data is then estimated to be 0.032~Jy, leading to a fluence ($\equiv S_i W$) of approximately 0.16~Jy~ms. This indicates that the high sensitivity of our data is capable of detecting low fluence FRBs as well as weak single pulses from RRATs. 

Following the above method, we estimated the single-pulse sensitivity for different telescopes, assuming a pulse width of 1~ms (see Fig.~\ref{fluence}). We note that the telescopes operate at different frequencies in general. Therefore, we assumed a flat spectral index in the calculation \citep[see Section 7.5 in][for discussion]{aab+21}. This estimation shows that the Arecibo telescope was capable of detecting all the known FRBs to-date and its sensitivity was second only to the 500-m FAST telescope.

\section{Single-pulse search}
\label{pipeline}

Since we are interested in searching for fast transients, we performed a standard single-pulse search analysis. We first determined the DM range to search for  according to our observation configuration. The dispersion smearing across the full bandwidth of our data is estimated to be $1.06\times$DM~ms, where DM is in pc~cm$^{-3}$ \citep[using Equation~4.7 in][]{lk12}. Therefore, the maximum DM that can be searched for in our data is $\sim$11\,000~pc~cm$^{-3}$ in order to ensure that the entire dispersed signal covers the bandwidth within the sky transit time of $\sim$13~s, optimizing the S/N. The signal would only partially cover the bandwidth for DMs greater than this value thus degrading the S/N. According to the relationship between the DM due to intergalactic plasma and the redshift \citep[using Equation~2 in][]{iok03}, this particular DM limit is equivalent to a redshift of $\sim$14.5, for $\Omega_m = 0.3$ and $\Omega_\Lambda = 0.7$ \citep[see][]{cp05,kdn+09}. 

In keeping with standard procedure of transient searches, we first dedispersed the data of each beam with trial DMs in the range of 1$-$11\,000~pc~cm$^{-3}$, excluding obvious local RFI with zero DM. The dedispersed data were then averaged in frequency to generate a time series and then searched for single pulses. In order to enhance the S/N of single pulses, the time series was convolved with a series of box-car filters with various widths between 1 and 4096 samples (i.e. between $\sim$0.0655--270~ms in time according to our $\sim$65.5~$\mu$s sampling time) in power-of-two increments. The information of the selected single pulses (e.g. time stamp, DM, box-car width) above a S/N threshold of 6 were saved for further evaluation. All the above steps were completed using the single-pulse search software package \textsc{heimdall}\footnote{\url{https://sourceforge.net/projects/heimdall-astro}}, which processes the data in parallel using GPUs to speed up the incoherent dedispersion \citep{bbbf12}. We also note that background continuum sources mentioned in Section~\ref{stability} were removed by subtracting running averages of two seconds from the time series before they were searched for single pulses. 

Single-pulse searches in general produce large number of candidates and therefore, it is not feasible to inspect all of them visually. For example, the \textsc{heimdall} package detected several thousand candidates for a given beam for each hour of our data, resulting in more than 80\,000 candidates for all seven beams per day. We also note that these candidates include a high rate of false positives due to Gaussian noise and RFI in the data. 
Therefore, machine learning techniques and algorithms have been introduced in single-pulse and pulsar searches to classify the candidates (e.g., RFI/non-astrophysical, FRB, and pulsar candidates). For instance, the deep learning methods have been used in pulsar search campaigns and improved the efficiency of searching procedures significantly \citep[see][]{zbm+14, dgm16, gdw+17, bd18}. Machine learning techniques have also been used in FRB searches \citep{wtt+16, fkg+18}, including convolution neural network classification methods \citep{zgf+18,cv18}. After sorting the candidates using classifier models, the high-probability ones can then be visually inspected in order to determine whether they are real astrophysical signals.  

We fed all the candidates through the FRB search software \textsc{fetch}\footnote{\url{https://github.com/devanshkv/fetch}}, which uses deep neural networks for classification of FRBs, RRATs, and RFI \citep{aab+20a,aa20}. \textsc{fetch} uses single-pulse candidate information from \textsc{heimdall} as described above and produces frequency-time, DM-time images as well as the frequency averaged time series. The convolutional neural networks are applied on these images in the classification process. The software currently includes 11 deep learning models, which were trained using known and simulated FRB, pulsar, and real RFI signals. \citet{aab+20a} reported that these models have an accuracy above 99.5\% on their test data set which consists of real RFI and pulsar candidates. We note that \textsc{fetch} runs on GPUs, which accelerates the image creation of candidates and the classification significantly by processing several candidates in parallel. In this process, each candidate was assigned a probability from 11 separate classifier models. We first sorted and inspected the candidates based on their probabilities produced by the classifier {\it Model a}, which has the highest prediction accuracy of 99.8\% \citep[see][]{aab+20a}. We also summed up the probabilities from all 11 classification models to obtain the {\it score} (out of 11) for each candidate and proceeded to sort and visually inspect each one. 

We note that both \textsc{fetch} and \textsc{heimdall} have been used in many searches and have also been integrated into transient search pipelines \citep[e.g.][]{als+20,lbp+20}. As an independent test, we applied our single-pulse search pipeline on the raw data for RRATs J1908$+$13 and J1924$+$10 that were recently discovered by the PALFA collaboration\footnote{\url{http://www.naic.edu/~palfa/newpulsars}}. Our pipeline re-detected these sources successfully with high maximum single-pulse S/N values of 34 and 14 for J1908$+$13 and J1924$+$10, respectively. These S/N ratios are slightly higher than the discovery-reported S/N values based on the \textsc{presto}-based (\texttt{single\_pulse\_search.py}) pipeline \citep{rem02,kjr+11,ran11}. We note that it is known that the two methods can report slightly different S/N ratios based on their data processing frameworks; see \citet{aal+21} and \citet{gff+21} for a detailed discussion.

\section{Single-pulse candidate results}
\label{results}

\begin{figure}
\includegraphics[width=8.5cm]{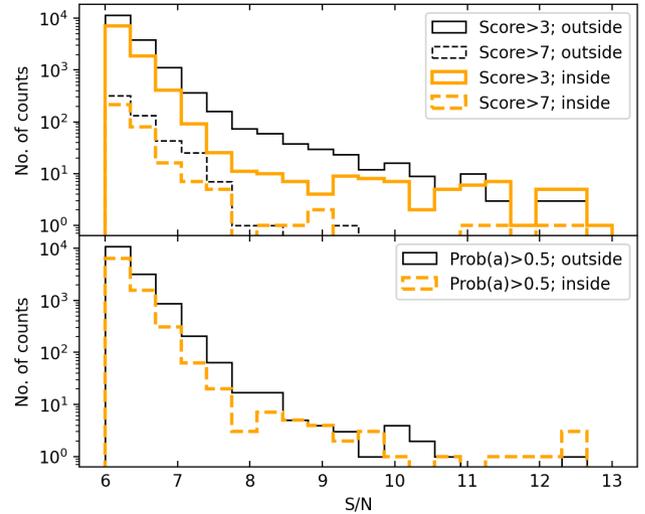}
\caption{The S/N distribution of single-pulse candidates that are placed outside ({\it black}) or inside ({\it orange}) the Galaxy using their DMs based on the electron density model YMW16. The {\it top} panel shows the distribution of candidates with {\it score}$>$3 ({\it solid}) and $>$7 ({\it dashed}). The bottom panel shows the distribution for candidates with probabilities $>$0.5 according to the classification {\it Model a}. All the candidates were visually inspected, but no potential FRBs or RRATs were identified.}
\label{hist_snr}
\end{figure}

The data processing produced more than $1.8\times10^6$ single-pulse candidates with S/N$>$ 6, but we reiterate that the false positive rate of these candidates is very high. We first sorted the candidates based on the probability of {\it Model~a} being greater than $0.5$. This criterion decreased the number of candidates to  23\,771, including only 563 with S/N$>$7. Out of these 563, the DMs of 401 candidates indicated that they are located beyond the Galaxy according to the YMW16 model. The visual inspection of these sorted candidates concluded that there are no potential FRB detections. We also used the {\it score}, which is the sum of probabilities determined from all classifier models (see Section~\ref{pipeline}) to sort all our candidates. The number of candidates with {\it score}$>$3 is 32\,254, and 5102 of those have S/N$>$7. Only 920 candidates with S/N$>$7  have high enough DMs to place them outside our Galaxy, and the visual inspection determined that there is no potential FRB detection. The S/N distribution of our candidates is shown in Fig.~\ref{hist_snr} for different $score$ values and probabilities of $Model~a$ thresholds. We also noticed that strong RFI signals on March 28 produced many high S/N single-pulse candidates that have DMs below 20~pc~cm$^{-3}$ in the search results. The classification models in \textsc{fetch}, however, did not identify them as RFI and provided high probabilities. We discarded them only after visual inspection. We also ignored the candidates that have the same DM and are appeared simultaneously in multiple beams as they are highly like to be due to non-astrophysical signals. We also sorted out all candidates with Galactic DMs based on the probabilities of {\it Model a} and also the {\it score} (see Fig.~\ref{hist_snr}). We visually inspected these candidates, but none of them showed any evidence of being a credible pulse from a RRAT or a pulsar.

\begin{figure*}
\includegraphics[width=17.5cm]{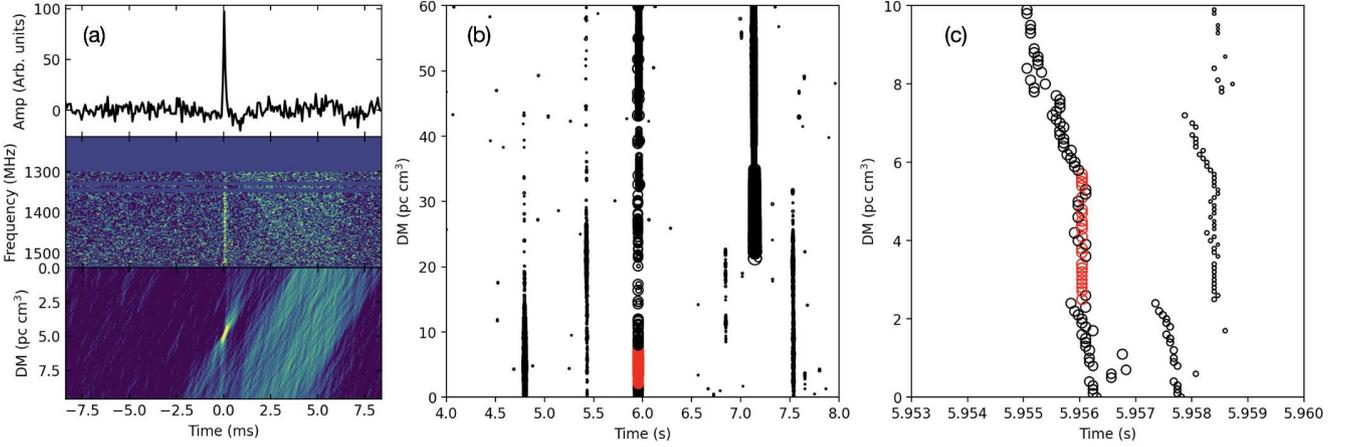}
\caption{High S/N ($\approx$17) candidate ({\it left} panel) in Beam4 on March 22 with a low DM ($\approx$4.8~pc~cm$^3$). The {\it left-middle} panel shows the candidate across the frequency band and the {\it left-top} panel shows the time series after averaging the signal across the frequency band. The {\it left-bottom} panel shows a color map of the pulse S/N across the DM-time space around its best reported DM value of 4.8~pc~cm$^3$. The candidate is independently validated using the single-pulse routine in the \textsc{presto} package and the DM versus time results are shown in panel (b) and (c). The circles represent a single-pulse detection at that particular DM and time, and the radius of the circle is scaled according to the detected S/N. The candidate on panel (a) corresponds to a signal around the time of $\sim$5.9~s on panel (b), which is shown in {\it red} circles. Panel (c) shows a zoomed-in version of panel (b) around the same candidate, indicating that the signal is not well constrained in DM and the candidate is likely due to RFI. }
\label{rfi_cand}
\end{figure*}

The classifier models evaluate the candidates mainly based on their training data sets and thus, there is a possibility that they can produce false results \citep[see][]{aab+20a}. Therefore, we visually inspected candidates that have detection S/N$>$7 regardless of their classifier-model-produced probabilities, but we still did not find any evidence of a transient detection. We note there were a few candidates with a reasonably high S/N detection; however, further examination indicated that these are mainly due to RFI or random noise. For example, Fig.~\ref{rfi_cand}(a) shows a \textsc{fetch}-produced plot for a high S/N ($\approx$17), low DM candidate. By processing the data around the same candidate independently with the standard single-pulse routine (\texttt{single\_pulse\_search.py}) in \textsc{presto} \citep{ran11}, we found that it is very likely due to RFI (see Fig.~\ref{rfi_cand}(b) and (c)). We also noticed a series of single pulses around the same candidate (with DM$<$70~pc~cm$^{-3}$) with slightly different time stamps. Therefore, we performed a periodicity search using \textsc{presto} to test whether this emission is from a pulsar, but the analysis confirmed that this signal is local and due to RFI.  

Furthermore, we searched for candidates that have the same sky location (within the beam width) and similar DM to identify repeaters. Such source can effectively identify due to its nature of repeating pulses and thus, we used a lower detection threshold of S/N $=$ 6 in the search. This analysis narrowed down a few sets of matching candidates, but visual inspection confirmed that they are likely due to noise and showed no evidence for detection of a repeater.

The major concern of candidate selection is how to distinguish real astrophysical signals and random noise when the candidate detection S/N is low (e.g. $\lessapprox$8). Most of our candidates fall into this S/N region (see Fig.~\ref{hist_snr}). \citet{yzw+21} recently presented a sample of example FRB candidates with very low S/N ratios using Parkes telescope archival data. The only way to prove that these weak events are astrophysical by re-detecting them in follow-up observations in the future. In order to understand the nature of candidates produced by random noise in our data, we searched for single-pulse candidates over `negative' DM values, which are purely `non-astrophysical' candidates produced by random noise. We selected for this purpose the $\sim$9~hr data set of Beam0 on March 25, which produced 9003 candidates with S/N$>$6 over the DM range of 1--11\,000~pc~cm$^{-3}$ in our standard search (described in Section~\ref{pipeline}). We fed the same data set into our pipeline with the same DM range, but for negative values (i.e., between $-$11\,000 to $-1$~pc~cm$^{-3}$). This analysis produced 10\,525 purely non-astrophysical candidates with S/N$>$6, which is more than the number of candidates found in the standard search. We also note that these non-astrophysical, random noise candidates are visually very similar to low S/N candidates in our standard search. This test suggests that low S/N candidates are not easily distinguished from real astrophysical signals and random noise. Thus, we conclude that the low S/N candidates in the distribution are due to random noise.

\section{Known pulsars and transients}
\label{known}

\begin{figure*}
\includegraphics[width=14cm]{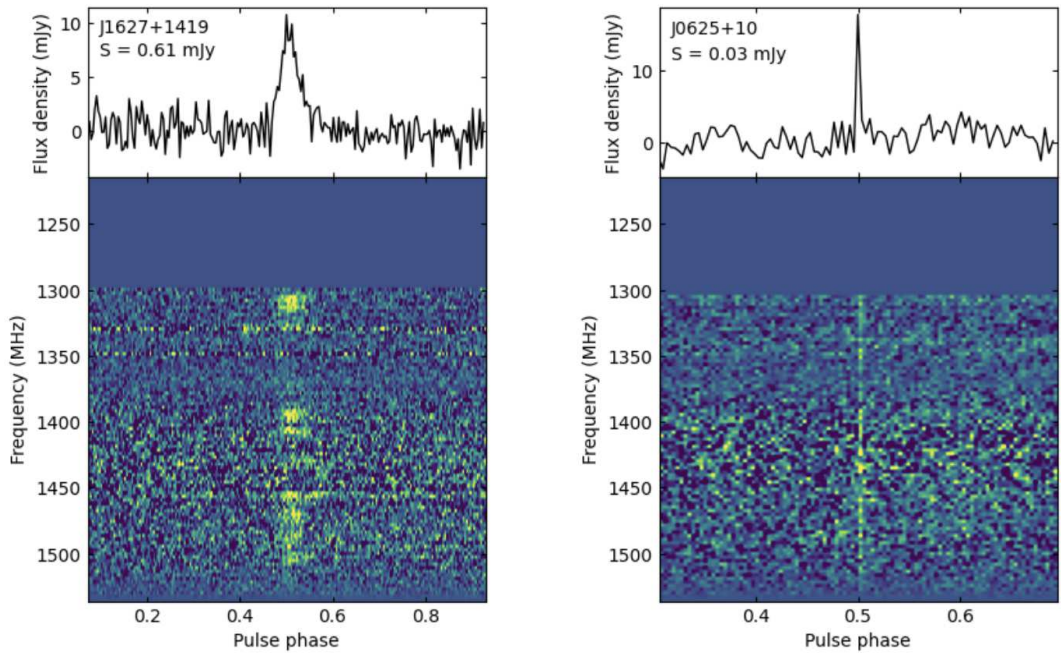}
\caption{Detected emission from known pulsars by folding the data using their timing ephemerides. PSR J1627$+$1419 was within Beam1 (with an offset of $0.5'$ from the beam center) on March 19, and J0625$+$10 was within Beam3 (with an offset of $1.4'$ from the beam center) on January 27. The {\it top} panel shows the integrated pulse profile, folded over 13~s of data, and the {\it bottom} panel shows the emission across the frequency band. The mean flux densities for J1627$+$1419 and J0625$+$10 are estimated to be $0.6\pm0.1$ and $0.03\pm0.1$~mJy, respectively. For clarity, only a portion of the pulse phase is shown.}
\label{known_psrs}
\end{figure*}

As shown in Fig.~\ref{sky}, there are five known pulsars that have positions covered by our observations. However, the single pulses of these sources were not detected through our pipeline. Perhaps these pulses were too weak to be detected with the noise level and our detection thresholds. In this section, we searched for their emission by folding the data using their timing ephemerides. The basic properties of these sources are given in Table~\ref{known_src}. PSR J1627$+$1419 was within our observations on March 19, and it was detected in Beam1 with an offset of $0.5'$ from the beam centre. This slow pulsar has a period of 0.491~s \citep{fcwa95} and it is bright at frequencies $<$800~MHz with fluxes of 78, 6, and 4 mJy at 150, 430, and 774 MHz, respectively, resulting in a spectral index of $-1.6\pm0.3$ \citep{bkk+16,lwf+04,hdvl09}. We extracted the $\sim$13~s data chunk in which the pulsar crossed Beam1 during its transit. We then processed this data using the ephemeris of the pulsar \citep{fcwa95} via the pulsar signal processing package \textsc{dspsr} \citep{vb11}. The pulsar emission is seen in the processed data, and the integrated pulse profile is shown in Fig.~\ref{known_psrs} (see left panel). We then estimated the flux density of the pulsar using the radiometer noise given in Equation~\ref{radio}. Multiplying this equation by $1/G$, where $G$ is the telescope gain, we can estimate the noise fluctuation in Jy \citep[see Equation 7.12 in][]{lk12}. Assuming $\Delta f = 215$~MHz, $t_{\rm obs}=13$~s, and $G=7.1$~K~Jy$^{-1}$ (see Section~\ref{stability}), as well as other standard values given in Appendix~\ref{gain}, we scaled the amplitude of the pulse profile in mJy (see Fig.~\ref{known_psrs}) and then estimated the mean flux density. The mean flux density we obtained is $0.6\pm0.1$~mJy, which is within the errors of the expected mean flux density of $0.9\pm0.4$~mJy at 1.4~GHz based on the spectral index of $-1.6\pm0.3$ \citep{bkk+16}.

\begin{table*}
\begin{center}
\caption{The properties of known pulsars and FRBs that have positions covered by our observations. The previously reported mean flux densities $S$ at 1.4~GHz and fluences at 600~MHz of these pulsars and FRBs are given in sixth and seventh columns, respectively. The last column noted the detectability of these sources in our data. Note that none of these FRBs have been identified as repeaters.}
\label{known_src}
\begin{tabular}{lccccccc}
\hline
\multicolumn{1}{c}{Source} &
\multicolumn{1}{c}{Period} &
\multicolumn{1}{c}{DM} &
\multicolumn{1}{c}{$l$} &
\multicolumn{1}{c}{$b$} &
\multicolumn{1}{c}{$S$} &
\multicolumn{1}{c}{Fluence} &
\multicolumn{1}{c}{Detected} \\
\multicolumn{1}{c}{} &
\multicolumn{1}{c}{(ms)} &
\multicolumn{1}{c}{(pc~cm$^{-3}$)} &
\multicolumn{1}{c}{($^\circ$)} &
\multicolumn{1}{c}{($^\circ$)} &
\multicolumn{1}{c}{(mJy)} &
\multicolumn{1}{c}{(Jy~ms)} &
\multicolumn{1}{c}{} \\

\hline
PSR~J0625$+$10 & 498 & 78 & 200.88 & $-$0.96 & 0.09 & -- & Yes \\
PSR~J1329$+$13 & -- & 12 & 338 & 73.99 & -- & -- & No \\
PSR~J1627$+$1419 & 491 & 32.2&  30.03 & 38.32 & 0.95$^\dagger$ & -- & Yes \\
PSR~J1651$+$14 & 828 & 48 & 32.88 & 33.07 & -- & -- & No \\
PSR~J1749+16 & 2312 & 59.6 & 41.21 & 20.90 & -- & -- & No \\
\hline
FRB20180906A & -- & 383.46 & 217.17 & 40.61 & -- & 3 & No \\
FRB20190131C & -- & 507.76 & 241.74 & 60.05 & -- & 2 & No \\
FRB20190214C & -- & 533.11 & 20.45 & 64.93 & -- & 5 & No \\
FRB20190224C & -- & 497.4 & 203.48 & 27.2 & -- & 8 & No \\
\hline
\end{tabular}
\begin{tabular}{l}
$^\dagger$The expected flux density of the pulsar estimated at 1.4~GHz based on its measured spectral index of $-1.6$ given in \citet{bkk+16}.\\
\end{tabular}
\end{center}
\end{table*}

PSR~J0625$+$10 happened to be within Beam3 on January 27 ($\sim1.4'$ off from the beam center). This pulsar has a spin period of $\sim$0.5~s, DM of 78~pc~cm$^{-3}$ \citep{cnst96}, and a very low flux density of 0.09~mJy at 1.4~GHz \citep{lbh+15}; see Table~\ref{known_src}. We again selected a $\sim$13~s data section during which the pulsar was within Beam3 and processed it using the ephemeris of the pulsar. The processed data is shown in Fig.~\ref{known_psrs} (right panel), and the observed pulse profile has a S/N of 8. The mean flux density is estimated to be $0.03\pm0.10$~mJy, following the same method given above, which is within the errors of the previously reported value \citep[see Table~\ref{known_psrs} and also][]{lbh+15}.

We also noticed that PSR J1329$+$13 was within Beam0 (offset by $0.5'$ from the beam centre) during the March 17 session based on the pulsar position reported in \citet{ttm+18}. This is a RRAT with a DM of $12\pm2$~pc~cm$^{-3}$, and its spin period is currently unknown. To identify emitted single pulses from the pulsar in our single-pulse analysis (see Section~\ref{pipeline}), we searched over all our candidates to find those that matched the DM and position of the pulsar. We used a tolerance of $\pm2$~pc~cm$^{-3}$ in DM to narrow down any possible candidates and constrained the position to be within our beam size. However, we did not find any convincing candidates that match the pulsar DM and position. PSR J1329$+$13 was discovered at a frequency of 111~MHz with a high flux density \citep{ttm+18}, suggesting that it should be detected in our data. That we did not detect this pulsar could imply that it was in the emission off state when it transited the beam. We further note that the current position of this pulsar is poorly constrained, RA~$=$~13$:$29(2) and Dec~$=$~$+$13$:$44(20)  
\citep{ttm+18}, in which case perhaps it was never actually within the beam.  

PSR J1749$+$16 is a 2.3~s pulsar with a DM of 59.6~pc~cm$^{-3}$ \citep{dsm+16}; see Table~\ref{known_src}. It was in Beam6 on March 23, and while we applied the same procedure as described above, we did not detect emission from the pulsar. We note that J1749$+$16 is a nulling pulsar with nulls for tens of seconds, suggesting that perhaps it was in the null state when it transited across the beam, resulting in non-detection. 

PSR J1651$+$14 is a 0.83~s pulsar with a DM of 48~pc~cm$^{-3}$ \citep{ttk+17}, and it was in Beam6 on March 19. As for other pulsars, we cropped the data section based on the reported position of the pulsar given in \citet{ttk+17} and then processed the data; however, we did not detect pulsar emission. We note that the discovery of the pulsar was very weak and carried out at 111~MHz. No other observations have been reported for this pulsar. Perhaps the pulsar is intrinsically weak and has a steep spectrum, so that its emission may not be visible at 1.4~GHz with our 13~s integration time. 

The FRBs given in Table~\ref{known_src} and also shown in Fig.~\ref{sky} also happened to be within the ALFA beams. In order to check for repeating emission from these FRBs, we selected the single-pulse candidates produced when these sources were within the beams and then searched it for candidates with DMs similar to these sources. However, we could not find any repeating emission from these FRBs in our analysis. We also note that these sources have not been reported as repeaters \citep[see][]{aab+21} and thus it is unlikely to expect detection of these sources in our data. 

Finally, magnetars can also produce FRB-like fast transient signals \citep[see][]{abb+20,brb+20,bbb+21}. Therefore, we looked for magnetars\footnote{\url{http://www.physics.mcgill.ca/~pulsar/magnetar/main.html}} that were within our observations based on their sky positions \citep{ok14,pag+21} in order to search for transient emission from them. However, none of these sources were within our beams.

\section{Discussion}
\label{dis}

We collected 160~hrs of drift-scan data over 23~days in January and March 2020 with the Arecibo telescope. We processed the data and searched for fast transients, FRBs and RRATs, using a single-pulse pipeline that includes the \textsc{heimdall} and \textsc{fetch} packages. The pipeline produced over $8\times10^6$ single-pulse candidates, and the neural networks classification models in \textsc{fetch} reduced this number to $\sim$24\,000 ``good'' candidates with probabilities $>$0.5. Out of the remaining candidates, only $<$1000 had S/N$>$7 (see Section~\ref{results}). We proceeded to inspect these candidates manually, but we did not identify any transient detections. We also searched over all of the candidates to find repeating transients by matching their DMs and sky locations, but again we found no evidence of detection for these sources. While there were no transient detections, we did observe emission from two known pulsars (PSRs J1627$+$1419 and J0625$+$10), and their measured flux densities are consistent with the expected values (see Fig.~\ref{known_psrs} and Section~\ref{known}).

We finally estimated limits on the FRB event rates based on our observations, and compared that with other published rate estimates. Given a beam FoV of $0.022$~deg$^2$ and $\sim$160 hrs of observations in our campaign, we estimate the upper limit of the FRB event rate as $<$2.8$\times10^5$~sky$^{-1}$~d$^{-1}$ at 1.4~GHz above a fluence of 0.16~Jy~ms. We also estimated the FRB rate using the published rates based on the detections from other telescopes. Since the Parkes telescope detected FRBs at 1.4~GHz and our observations were also at the same frequency band, we used the Parkes FRB rate to estimate the expected FRB rate for our observations. \citet{tsb+13} reported FRB detections using Parkes and estimated the rate as $10\,000$~sky$^{-1}$~d$^{-1}$. \citet{kp15} reanalyzed \citet{tsb+13} results and derived a fluence complete event rate of 2500~sky$^{-1}$~d$^{-1}$ above a fluence of 2~Jy~ms. As described in Section~\ref{sensitivity}, the sensitivity of our observations is estimated at a fluence of 0.16~Jy~ms assuming a pulse width of 5~ms. In order to convert the sensitivity of Parkes to Arecibo, we use the FRB power-law flux distribution $N(>F) \propto F^{-\gamma}$, assuming a Euclidean Universe with $\gamma=3/2$, where the sources are assumed to be non-evolving and uniformly distributed in space \citep{clm+16}. We also note that $\gamma$ is constrained using the FRBs detected with the CHIME telescope to be $1.4$ \citep{aab+21}, which is an excellent match to an Euclidean space source distribution. Using this flux distribution for the Parkes event rate with its own fluence given above, we scaled the FRB rate to 1.1$\times10^5$~sky$^{-1}$~d$^{-1}$ above our fluence limit of 0.16~Jy~ms. The recent FRB detections reported by CHIME constrained the event rate to be 818~sky$^{-1}$~d$^{-1}$ above a fluence of 5~Jy~ms at 600~MHz \citep{aab+21}. Using the above mentioned flux distribution with a flat spectral index, we then scaled the event rate to be $1.4\times10^5$~sky$^{-1}$~d$^{-1}$ above our fluence limit of 0.16~Jy~ms at 1.4~GHz. The spectral properties of FRBs are still poorly understood and thus, a flat distribution is a valid assumption \citep[see][for discussion]{msb+19,ffb+19,aab+21}.

Finally, we estimated the average observing time required for an Arecibo-like telescope to detect at least one FRB using the event rate of 1.1$\times10^5$~sky$^{-1}$~d$^{-1}$ estimated above. Given the ALFA beam FoV of 0.022~deg$^2$, the average required observation time to detect one FRB is approximately 410 hours. Assuming eight hours of continuous observations every day, we need, on average, at least 51 days of observations, which is a factor of 2.2 longer than we spent in our campaign. Note that we conducted these observations during the observatory closure to fill the downtime of the telescope, so that we could not observe more than 160 hours to satisfy the above detection requirement.

Currently, we are processing our data to search for pulsars using periodicity searches. Even though the sky transit time within our beams is $\sim$13 ~s, we emphasize that it is worth searching for pulsars in our data, and bright pulsars, especially those with short periods, may appear with a reasonable S/N ratio (see the known pulsar detections in our data that described in Section~\ref{known_psrs}). In parallel to the periodicity search, we have included \textsc{presto}-based \citep{rem02,kjr+11,ran11} single-pulse routines (\texttt{single\_pulse\_search.py}\footnote{\url{https://github.com/scottransom/presto}}) in our pipeline to conduct an independent transients search analysis. The data processing is underway, and we will relay those results in a future publication.

We conclude by noting that our transient search pipeline is currently under modification to perform real-time searches, and it will be tested on the 12-m telescope at the Arecibo Observatory. This telescope is undergoing an upgrade to integrate a cooled receiver system\footnote{\url{http://www.naic.edu/~phil/hardware/12meter/patriot12meter.html}}$^,$\footnote{\url{http://www.naic.edu/ao/scientist-user-portal/astronomy/under-development}}. A significant portion of the telescope time will be dedicated to real-time transient searches, along with commensal observations, in the future.

\section*{Data availability}

The data underlying this article will be shared on reasonable request to the corresponding author.

\section*{Acknowledgments}
The Arecibo Observatory is operated by the University of Central Florida, Ana G. Mendez-Universidad Metropolitana, and Yang Enterprises under a cooperative agreement with the National Science Foundation (NSF; AST-1744119). We thank Emilie Parent for useful discussions about some of the data processing tools.


\bibliography{psrrefs,modrefs,journals}
\bibliographystyle{mnras}


\appendix

\section{Telescope gain and stability estimation}
\label{gain}

The telescope gain was obtained using the time series corresponding to a spectral channel from the observations during the transit of a continuum source. The spectral channel was selected in a way such that the data is not affected by RFI. Since the expected source transit time within our beams is $\sim$13~s, we average the raw data to make 1~s integration intervals. The NVSS source catalog was used to identify the sources and obtain their flux densities \citep{ccg+98}. The sources with flux densities greater than 100~mJy at 1.4~GHz and those fall within $1'$ of the beam were selected. This criteria resulted in at least one source in each beam almost everyday. This allows us to estimate the gain every day for each beam, determining the stability of the system. The gains were estimated using two methods.

The first method uses the radiometer equation, where the expected root mean square noise fluctuation 
\begin{equation}
\label{radio}
\Delta T_{\rm sys} = \frac{T_{\rm sys}}{\sqrt{n_{\rm p} t_{\rm obs} \Delta f}} = C \sigma_{\rm p}
\end{equation}
where $T_{\rm sys}$ is the system temperature, $n_{\rm p}$ is the number of polarization channels summed, $t_{\rm obs}$ is integration time, $\Delta f$ is observing bandwidth, and $\sigma_{\rm p}$ is the standard deviation of the off-source region in the time series. The proportionality constant $C$ is in units of K per count and used to scale the ordinate of the time series. In the calculation, we used the typical value $T_{\rm sys}=28$~K of ALFA receiver, $\Delta f = 0.335$~MHz, $n_{\rm p}=2$, and $t_{\rm obs}=1$~s. The constant $C$ is then used to convert the on-source deflection in counts to units of K. The telescope gain $G$ is the ratio of the deflection in units of K to the flux density of the source in Jy. The direct estimate of $\sigma_{\rm p}$ from the time series may affect by the gain variation over time and confusion noise. Therefore, we computed $\sigma_{\rm p}$ value from the difference between the time series from adjacent channels, leading to cancel out both these effects.

In the second method, we estimated the gain using the median of the off-source region of the time series, which is  proportional to $T_{\rm sys}$. Thus, the ratio of $T_{\rm sys}$ to median value gives the conversion factor in units of K per count. The on-source deflection can then be converted to K using this factor and the telescope gain can be estimated as described above.

The gain estimated using these two methods are found to be comparable with each other. Therefore, we averaged the gain obtained from the two methods for each beam. We note that one of the polarization channels of Beam5 was unstable and poorly behaved, so that it was ignored in the gain estimation.


\bsp	
\label{lastpage}
\end{document}